\documentclass[aps,prl,twocolumn,showpacs,showkeys]{revtex4-1}

\usepackage{amssymb}
\usepackage{amsmath}

\begin{document}


\title{A Process Algebra Approach to Quantum Field Theory}


\author{William H. Sulis}
\affiliation{McMaster University}
\email{sulisw@mcmaster.ca}

\date{\today}

\begin{abstract}
The process algebra has been used  successfully to provide a novel formulation of quantum mechanics in which non-relativistic quantum mechanics (NRQM) emerges as an effective theory asymptotically. The process algebra is applied here to the formulation of quantum field theory. The resulting QFT is intuitive, free from divergences and eliminates the distinction between particle, field and wave. There is a finite, discrete emergent space-time on which arise emergent entities which transfer information like discrete waves and interact with measurement processes like particles. The need for second quantization is eliminated and the particle and field theories rest on a common foundation, clarifying and simplifying the relationship between the two.
\end{abstract}


\maketitle
\section{Introduction}

The Process Algebra program seeks to develop a theory of fundamental phenomena which is a true completion of quantum mechanics and not merely a reformulation \cite{Sulis-th, Sulis-jmp, Sulis-arxiv}. It proposes a discrete, finite but \emph{generated} space-time in which fundamental entities such as particles and fields appear as \emph{emergent} phenomena. It avoids the form of non-locality that plagues quantum mechanics and conflicts with special relativity. It retains a limited form of non-contextuality. Measurement is viewed as an interaction between a process and a specialized measurement process without requiring a separate theory. The simple shift from a continuous, global, eternal perspective to a discrete, finite, generated perspective males all the difference. The process approach appears to resolve many of the paradoxes of quantum mechanics and yet remains intuitive. This approach has been applied with some success to the case of scalar (and by extension vector) particles in non-relativistic quantum mechanics \cite{Sulis-th}. Here this theory is extended to quantum fields. A relativistic extension taking into account symmetries and  interactions will be presented elsewhere \cite{Edvinsson}. Some basic ideas of the process algebra are given in the appendix.

\section{Process Approach to the Single Scalar Particle}

According to  the process approach, a process generates primitive elements termed \emph{informons}, which are organized into a discrete, finite, causal antichain termed a \emph{causal tapestry}. A causal tapestry may be thought of as consisting of discrete, causally unrelated elements of space-time, each possessing a local contribution to an emergent particle or field. From the current tapestry $\mathcal{I}$, a process $\mathbb{P}$  generates a new tapestry $\mathcal{I}'$ in a series of rounds (see Appendix), following which the procedure repeats using $\mathbb{P}$ (or a new process $\mathbb{P}'$ if interactions have taken place). The causal tapestry is thought of as \textquotedblleft reality\textquotedblright and the standard models as \textquotedblleft ideals \textquotedblright\ so to compare the two, each informon must be \emph{interpreted} \cite{Sulis-ad} in terms of the mathematics of a standard model. Thus each causal tapestry is associated with a causal manifold \cite{Borchers} $\mathcal{M}$ (causal distance $d$) and a Hilbert space $\mathcal{H}(\mathcal{M})$ on $\mathcal{M}$. Each informon is described as $[n]<\mathbf{m}_{n},\phi_{n}(\mathbf{z}),\Gamma_{n},\mathbf{p}_{n}>\{G_{n}\}$ where $\mathbf{m}_{n}\in \mathcal{M}$ (interpretation as a space-time point), $\phi_{n}(\mathbf{z})\in \mathcal{H}(\mathcal{M})$ (interpretation as a local interpolation contribution to a wave function), $\Gamma_{n}$ (local strength of the generating process), $\mathbf{p}_{n}$ (set of local properties inherited from the generating process). $G_{n}$ is a causally ordered set of informons from previously generated but now defunct causal tapestries (interpretation as the content or information  of $n$). The set $\cup _{n\in \mathcal{I}}G_{n}$ forms a coherent causally ordered, hence  the name \emph{causal} tapestry.  There is also a causal distance function $\rho$ which is space-like on $\mathcal{I}$ and time-like or null on the edges of $G_{n}$ and $\rho(n,m)=d(\mathbf{m}_{n},\mathbf{m}_{m})$. The informons of the causal tapestry form a causal antichain. The information that enters into the construction of an informon $n$ is wholly contained within $G_{n}$, thus only transmitted causally and \emph{never} in violation of special relativity.

Each local Hilbert space contribution has the form $\phi_{\mathbf{m}}=\Gamma_{\mathbf{m}}T_{\mathbf{m}}g(\mathbf{z})$ where $g(\mathbf{z})$ is an interpolation kernel on $\mathcal{M}$ and $T_{\mathbf{m}}$ is a translation operator (eg. in one dimension $T_{x}g(z)=g(z-x)$). The
process theory of measurement \cite{Sulis-th,Sulis-arxiv} posits that the Born rule is emergent. All relevant probabilities may be calculated solely based upon the values $\{\Gamma_{n}, \mathbf{p}_{n}, n\in \mathcal{I}\}$. Thus \emph{all} of the relevant physics is determined on the informons alone, independent of the interpretations.

A single scalar particle is represented as a discrete set of informons, interpreted as a discrete set of points in the causal manifold $\mathcal{M}$ and as a physical wave via the global Hilbert space $\mathcal{H}(\mathcal{M})$  interpretation $\Phi(\mathbf{z})=\sum_{n\in \mathcal{I}}\Gamma_{n}T_{\mathbf{m}_{n}}g(\mathbf{z})$ where the local value of the wave is the local strength of the generating  process.

A simplified process algebra model has been applied to the case of scalar (and by extension vector) particles in NRQM where $\mathcal{M}=\mathbb{R}^{4}$ and the causal tapestry $\mathcal{I}_{n}$ embeds into $\{n\}\times \mathbb{R}^{3}$ as a regular lattice with spacing $l_{P}$ (Planck length) and $g$ is taken to be a 4-d sinc function \cite{}. The local strengths in $\mathcal{I}_{n+1}$ are generated from $\mathcal{I}_{n}$ by propagating information forward using  the Green's function $K(\mathbf{x},\mathbf{y})$ for the corresponding Schr\"odinger's equation (which must depend only on $d (\rho))$. If  for each $m\in\mathcal{I}_{n+1}$, $\mathcal{I}^{m}_{n}$ is the set of informons of $\mathcal{I}_{n}$ propagating information forward to $m$, then $\Gamma_{m}=\sum_{k\in\mathcal{I}^{m}_{n}} l_{P}^{3}K(\mathbf{m}_{m},\mathbf{m}_{k})\Gamma_{k}$. Let $\hat c$ denote the numerical value of the speed of light $c$, sans units. In this simple model, each informon $n$ receives information from prior informons lying within a ball of diameter $\hat c l_{P}$ centered on the spatial component of $\mathbf{m}_{n}$ and conversely, $n$ can contribute information to create informons lying within a similar ball in $\mathcal{I}_{n+2}$. This corresponds to setting a maximal speed of information transfer of $c$.  In the asymptotic limit of $l_{P}\rightarrow 0$ the global Hilbert space interpretation converges to the usual Schr\"odinger wave function, and NRQM is seen as an effective or emergent theory \cite{Sulis-th,Sulis-jmp}.

\section{Process Approach to Fields}
Classically, particles and fields are considered to be distinct classes of physical entities both phenomenologically and in their mathematical representations. This distinction carried through in quantum mechanics with notions of quantization and Hilbert spaces for particles and second quantization and Fock spaces for fields. The process algebra approach is much simpler: a process is considered to be particle-like when few informons are generated per round and field-like when large numbers of particles are generated per round. In the latter case, simultaneous measurements at many spatially separated sites will yield results with high probability, thus satisfying Feynman's notion of a field as \textquotedblleft ... a set of numbers we specify in such a way that what happens \emph{at a point} depends only on the numbers \emph{at that point}" (sic) \cite{Mead}.
A quantum field is an idealization in the continuum limit as $R\rightarrow \infty$. In the process algebra model there is no wave-particle duality. Every physical entity has discrete (interpreted via $\mathcal{M}$) and continuous (interpreted via $\mathcal{H}(\mathcal{M})$) aspects. There also is no intrinsic distinction between particle and field,  both having representation within the \emph{same} process algebra. The picture is intuitive and mathematically simpler.

Let $P$ denote a collection of primitive processes representing the distinct pure states of a single entity. Let $\Pi_{P}$ denoted the process algebra generated by $P$ and let $\Sigma_{P}$ denote the subalgebra of $\Pi_{P}$ formed by taking all $\oplus,\hat\oplus$ sums of elements of $P$. The space of bosonic-like fields is defined as the subalgebra of $\Pi_{P}$ of the form $\Sigma_{P}\oplus (\Sigma_{P}\hat\otimes\Sigma_{P})\oplus(\Sigma_{P}\hat\otimes\Sigma_{P}\hat\otimes\Sigma_{P})\oplus\cdots$ while the space of fermionic-like fields is defined as the subalgebra of $\Pi_{P}$ of the form
$\Sigma_{P}\oplus (\Sigma_{P}\otimes\Sigma_{P})\oplus(\Sigma_{P}\otimes\Sigma_{P}\otimes\Sigma_{P})\oplus\cdots$.
These are the process algebra analogues of the Fock space. Note that unlike the Fock space, an element of either of these subalgebras is just  a complicated process, not a wholly new  mathematical form. A quantum field arises when the number of terms can be taken to be countable infinite, for example $\mathbb{P}\otimes \mathbb{P}\otimes\mathbb{P}\otimes\cdots$.

 \section{PCM for  single particles}
 
Reality according to the process model is discrete and finite. NRQM is an idealization which applies whenever a continuum approximation can be considered. The distinction between the two can be made evident through a \emph{set valued} map \cite{Aubin}, the process covering map (PCM).  An  active primitive process $\mathbb{P}$ acting on a tapestry $\mathcal{I}$ generates a sequence  (corresponding to the succession of rounds) of partial causal tapestries $\emptyset,\mathcal{I}_{1}',\mathcal{I}_{2}',\mathcal{I}_{3}'\ldots$, each formed from the previous tapestry by the inclusion of an informon, $\mathcal{I}_{i}'=\mathcal{I}_{i-1}'\cup\{n_{i}\}$. The sequence of partial tapestries forms an ordered set with edges labelled by informons $n_{1}, n_{2}, \ldots, n_{k}, \ldots$($\mathcal{I}_{i-1}'\stackrel{n_{i}}{\rightarrow}\mathcal{I}_{i}'$) and having a maximal element, the final causal tapestry.  Letting $\mathbb{P}$ act upon $\mathcal{I}$ again will generate a different  ordered set of tapestries with edges $n_{1}', n_{2}', \ldots, n_{k}', \ldots$. Two distinct global Hilbert space interpretations will be generated $$\Phi^{1}(\mathbf{z})=\sum_{n_{i}} \phi_{n_{i}}(\mathbf{z})\text{ and } \Phi^{2}(\mathbf{z})=\sum_{n_{i}'} \phi_{n_{i}'}(\mathbf{z})$$

The union of all possible tapestry sequences forms the \emph{process sequence tree} of $\mathbb{P}$ with initial tapestry $\mathcal{I}$, denoted $\Sigma(\mathbb{P},\mathcal{I})$.  Associate to  $\mathbb{P}$ a  set $H_{\mathbb{P}}$ of elements of $\mathcal{H}(\mathcal{M})$ consisting of all global $\mathcal{H}(\mathcal{M})$-interpretations constructed from every maximal tapestry in the sequence tree.  
In the limit $N,r\rightarrow \infty$ (see Appendix) interpolation theory shows that  $H_{\mathbb{P}}\rightarrow \{\Phi^{l_{P}}(\mathbf{z})\}$, a singleton set, which will correspond to the NRQM wave function under the asymptotic condition $l_{P}\rightarrow 0 $, provided that propagation is via a suitable discrete Schr\"odinger Green's function.

The process algebra model is thus seen as a dynamical completion of NRQM, providing information about the actual generation of informons corresponding to physical entities. The theory of measurement within the process model shows that the usual statistical structure of NRQM is obtained under suitable asymptotic limits \cite{Sulis-th}. The departure from NRQM predictions away from the asymptotic limit offers the opportunity for testing the process model.

For fixed $\mathcal{I}$, and some primitive process $\mathbb{P}$, define the PCM $\mathfrak{P}_{\mathcal{I}}:P\rightarrow \mathcal{H}(\mathcal{M})$  by $\mathfrak{P}_{\mathcal{I}}(\mathbb{P})=H_{\mathbb{P}}$. {Technically $\mathfrak{P}_{\mathcal{I}}(\mathbb{P})$ is a set-valued map so one should write $\mathfrak{P}_{\mathcal{I}}:P\rightarrow \mathcal{P}(\mathcal{H}(\mathcal{M}))$, where $\mathcal{P}(\mathcal{H}(\mathcal{M}))$ is the power set on $\mathcal{H}(\mathcal{M})$ but $\mathcal{H}(\mathcal{M})$ is simpler. By allowing $\mathcal{I}$ to vary one obtains an operator interpretation for the PCM \cite{Sulis-th}. 

Define the formal product $w\mathbb{P}$ to mean that the tokens generated by the actions of $\mathbb{P}$ have their value multiplied by $w$. Then a simple argument shows that $$\mathfrak{P}_{\mathcal{I}}(\oplus_{i} w_{i}\mathbb{P}_{i})=\sum_{i} w_{i}\mathfrak{P}_{\mathcal{I}}(\mathbb{P}_{i})=\mathfrak{P}_{\mathcal{I}}(\hat\oplus_{i} w_{i}\mathbb{P}_{i})$$ where for two sets of functions $A,B$ the sum $$A+B=\{f+g|f\in A,g\in B\}$$ which extends the map to $\Sigma_{P}$, the sum algebra over $P$.

\section{PCM and PCM${}^{C}$ for Fields}
Fields consist of products of primitive processes, requiring an extension of the PCM from $\Sigma_{P}$ to $\Pi_{P}$. The construction is similar for both free and exclusive products though with different  sets of Hilbert space interpretations. There are  two different kinds of PCM to consider, one  phenomenological and the other  correlational providing a  link to the usual quantum mechanical formulation. 

Consider a generic product $\mathbb{P}=\Pi_{i} \mathbb{P}_{i}$ of primitive processes $\mathbb{P}_{i}$. As above one may construct the process sequence tree $\Sigma(\mathcal{I},\mathbb{P})$ except that in this case at each round $i$ the product process will generate a correlated  \emph{set} of informons $A_{i}=\{n^{1}_{i},\ldots,n^{n}_{i}\}$ ($n^{k}_{i}$ generated by $\mathbb{P}_{k}$). The phenomenological approach uses the co-product construction. Partition the new causal tapestry $\mathcal{I}'$ into subsets of informons generated by the individual subprocesses, i.e. $\mathcal{I}'=\cup_{i}\mathcal{I}_{i}'$ ($\mathcal{I}_{i}'=\{n^{i}_{1},n_{2}^{i},\ldots\}$). The global Hilbert space interpretation $\Phi_{\mathcal{I}'}(\mathbf{z})$ is decomposed onto each of these component sets yielding $\Phi(\mathbf{z})_{i}'$ on $\mathcal{I}_{i}'$ (an element of $\mathfrak{P}_{\mathcal{I}}(\mathbb{P}_{i})$) and thus  given as the formal co-product sum  $\Phi_{\mathcal{I}'}(\mathbf{z})=\Phi_{1}'(\mathbf{z})\oplus \Phi_{2}'(\mathbf{z})\oplus \cdots \oplus \Phi_{j}'(\mathbf{z})$. Note that the functions are \emph{not} evaluated at a point.  Using the co-product one defines $\mathfrak{P}_{\mathcal{I}}(\otimes_{i} \mathbb{P}_{i})=\mathfrak{P}_{\mathcal{I}}(\mathbb{P}_{1})\sqcup\mathfrak{P}_{\mathcal{I}}(\mathbb{P}_{2})\sqcup\cdots\sqcup\mathfrak{P}_{\mathcal{I}}(\mathbb{P}_{n})$.

If the subprocesses represent different states of a single  process then the co-product can be replaced by sums.

The co-product describes multiple entities co-evolving, hence providing a phenomenological representation.  Quantum mechanics, however,  is concerned with correlations among measurements, which, in the process model, are triggered by the generation of individual informons,  so correlations depend not on the global Hilbert space interpretation but rather on the process sequence tree, which is a computational and combinatorial construction, not a phenomenological description. The analysis of correlations requires the construction of an artificial configuration space PCM (PCM). There are several subtleties which must be noted before proceeding.

The first subtlety is that if an informon $n^{k}_{j}$ is generated in round $j$ by $\mathbb{P}_{k}$, then it will never be generated in any other round. This implies that $\mathcal{I}'=\cup_{i}\mathcal{I}_{i}'\neq\prod_{i}\mathcal{I}_{i}'$ so that a single process action cannot generate all possible correlated sets of informons. A single tapestry $\mathcal{I}'$ must be expanded in order to determine process strengths for all possible informon collections generated by $\mathbb{P}$. Second, although the  informons within each set $A_{k}$ are correlated by virtue of the generating process $\mathbb{P}$, each $A_{k}$  and the informons within are generated independently. This implies that we may artificially extend the sequence process tree by taking a maximal causal set ${}^{1}\mathcal{I}'$ and extending it by an edge obtained from a different path so long as  we ensure that if we wish to add $n_{2}$ to component ${}^{1}\mathcal{I}_{i}'$ of ${}^{1}\mathcal{I}'$ and there exists in component ${}^{1}\mathcal{I}_{i}'$ of $\mathcal{I}_{1}$ an $n_{1}$ such that $\mathbf{p}_{n_{1}}=\mathbf{p}_{n_{2}}$ and  $\mathbf{m}_{n_{1}}=\mathbf{m}_{n_{2}}$ then $\Gamma_{n_{1}}=\Gamma_{n_{2}}$. Such an informon  is said to be admissible in ${}^{1}\mathcal{I}'$. Repeatedly extend the sequence tree by adding admissible elements to maximal tapestries until no further extensions are possible. The resulting tree is called the process configuration sequence tree $\Sigma^{C}(\mathcal{I},\mathbb{P})$. The maximal tapestries $\{{}^{1}_{M}\mathcal{I}',{}^{2}_{M}\mathcal{I}',\ldots\}$ of $\Sigma^{C}(\mathcal{I},\mathbb{P})$ have sufficient numbers of informon combinations to calculate correlations and so  it makes sense to define
the global configurational $\mathcal{H}(\mathcal{M})$ interpretation on one of these maximal tapestries as $\Phi^{C}_{{}^{i}_{M}\mathcal{I}'}(\mathbf{z})=$

$$\sum_{\{n^{1},\ldots,n^{n}\}\subset {}^{i}_{M}\mathcal{I}'}\Gamma_{n^{1}}\cdots\Gamma_{n^{n}}T_{\mathbf{m}_{n^{1}}}g(\mathbf{z}_{1})\cdots T_{\mathbf{m}_{n^{n}}}g(\mathbf{z}_{n})$$

\noindent where the sum is over the edge sets along the path forming ${}^{i}_{M}\mathcal{I}'$.

Given $\Sigma^{C}(\mathcal{I},\mathbb{P})$, let $\mathfrak{I}^{M}_{\Sigma^{C}(\mathcal{I},\mathbb{P})}$ denote the set of all of its maximal causal tapestries. We define the configuration process covering map, PCM${}^{C}$, denoted $\mathfrak{P}^{C}_{\mathcal{I}}(\mathbb{P})=$ to be

$$\{\sum_{(n^{1},\ldots,n^{n})\in\: {}^{i}_{M}\mathcal{I}'} \Gamma_{n^{1}}\cdots\Gamma_{n^{n}}T_{\mathbf{m}_{n^{1}}}g(\mathbf{z_{1}})\cdots T_{\mathbf{m}_{n^{n}}}g(\mathbf{z}_{n})\}$$

\noindent taken over all ${}^{i}_{M}\mathcal{I}'\in\mathfrak{I}^{M}_{\Sigma^{C}(\mathcal{I},\mathbb{P})}$. 
If no tuples $(n^{1},\ldots,n^{n})$ are excluded this may be written as

$$\{\sum_{n^{1}\in \mathcal{}^{j}_{M}{I}_{1}'}\Gamma_{n^{1}}T_{\mathbf{m}_{n^{1}}}g(\mathbf{z}_{1})\cdots\sum_{n_{n}\in {}^{j}_{M}\mathcal{I}_{n}'}\Gamma_{n_{n}}T_{\mathbf{m}_{n_{n}}}g(\mathbf{z}_{n})\}$$

\noindent resembling the more familiar configuration space construction.
Otherwise this factorization is not possible.
 An operator version of the PCM${}^{C}$ is under study.
\section{Conclusion}

The process algebra approach utilizes the same mathematical framework to describe both particles and fields, eliminating both wave-particle duality and particle/field distinctions. The resulting field theory appears intuitive and being discrete and finite, is inherently free of divergences and hopefully paradox as well. It is a generative theory and so has the potential to deal with open and non-stationary situations. If the Green's function used to generate informons is relativistically invariant then the process algebra model will be as well. Information transfers through causally local paths, and there is a limited form of non-contextuality with properties inherited from the generating processes (which are not complete due to the inherent non-commutativity of process concatenation). The configuration space is seen here to be purely heuristic, not phenomenological. More work is needed to understand the relation between the PCM${}^{C}$ and the usual operator formalism of QFT derived from second quantization. The process algebra version of the Klein-Gordon scalar field should closely approximate its quantum mechanical counterpart as it naturally extends the NRQM case. Different implementations of the process as combinatorial games should yield different global Hilbert space interpretations which should be testable against standard predictions and by experiment.
\section{Appendix: Process Algebra}

The notion of process, based on Whitehead \cite{Shimony} is modeled analogous to the use of combinatorial games \cite{Conway}  in logic \cite{Hirsch}. This was inspired by ideas of Rosen \cite{Rosen} and Trofimova \cite{Trofimova} to apply complex systems theory to fundamental phenomena. A process is a \emph{generator} of primitive events termed \emph{informons}, organized into a discrete and finite causal structure $\mathcal{I}$ termed a \emph{causal tapestry} \cite{Sulis-ct}. Through process, a  causal tapestry emerges and then fades away, becoming information which causally propagates to create the next tapestry \cite{Sulis-ad}.

 A process creates one or more informons in a series of short rounds $r$,  which together form a round, and then constructs the causal tapestry in a series of rounds $N$. A primitive process generates only one informon per round ($R=1$). Complex processes $(R>1)$ are formed from algebraic combinations of primitive processes. 

A process exists in one of two states: active or inactive. The subprocesses of a complex process may act concurrently, meaning that in a given round each subprocess will generate an informon (denoted as a product),  or they may act sequentially, meaning that in a given round only one subprocess generates an informon (denoted as a sum) (no fixed order of action is specified). All actions of process are held to be non-deterministic in the sense of computer science, logic and game theory \cite{Conway}. Subprocesses may contribute to the generation of the same informon (free) or only of distinct informons (exclusive). They may act independently or interact. These result in several different sums and products of processes: $\oplus$ (sequential, exclusive, independent), $\hat\oplus$ (sequential, free, independent), $\otimes$ (concurrent, exclusive, independent), $\hat\otimes$ (concurrent, free, independent), and corresponding interactive operations.
The zero process, $\mathbb{O}$, is the process that does nothing.

Processes may also be concatenated, which describes the outcome of interactions. including measurement (described elsewhere \cite{Sulis-th,Sulis-o}.  Note that sums and products are naturally Abelian, while concatenation is naturally non-Abelian.

As a general rule, processes corresponding to different states of a single entity combine through the exclusive sum while processes forming a single state of a single entity combine using the free sum. Boson-like processes corresponding to distinct entities may combine using either free or exclusive products while fermion-like processes corresponding to distinct entities combine using the exclusive product.


\begin{thebibliography}{aa}
\bibitem{Sulis-th} W. Sulis,   Ph.D. Thesis. \emph{A Process Model of Non-relativistic Quantum Mechanics}, University of Waterloo (2014)
\bibitem{Sulis-jmp} W. Sulis,  J. Mod. Phys. \textbf{5(6)} 1789 (2014).
\bibitem{Sulis-arxiv} W. Sulis,  arXiv: 1302-4156.
\bibitem{Edvinsson} U. Edvinsson, \emph{The Mammoth Book of Quantum Interpretation}.
(In preparation) (2015).
\bibitem{Sulis-ad} W. Sulis, NDPLS., \textbf{14(3)}, 209 (2010).
\bibitem{Borchers} H.J. Borchers and R.N. Sen,  \emph{Mathematical implications of Einstein-Weyl causality.} Springer-Verlag, New York (2006).
\bibitem{Mead} C. Mead, \emph{Collective Electrodynamics: Quantum Foundations of Electromagnetism}. MIT Press, Boston (2002).
\bibitem{Aubin} J.P. Aubin and H. Frankowska, \emph{Set valued analysis}. Birkhauser, Boston (2009).
\bibitem{Shimony} A. Shimony, \emph{Search for a Naturalistic World View, Volume II, Natural Science and Metaphysics.} Cambridge University Press, Cambridge, (1993).
\bibitem{Conway} J.H. Conway, \emph{On Numbers and Games}. A.K. Peters, Natick (2001).
\bibitem{Hirsch} R. Hirsch and I. Hodkinson, \emph{Relation Algebras by Games.} North-Holland, Amsterdam, (2002).
\bibitem{Rosen} R. Rosen, \emph{Quantum Implications: Essays in Honor of David Bohm.} B.J. Hiley \& F.D. Peat (eds) Routledge, London (1987).
\bibitem{Trofimova} I. Trofimova,  in:  \emph{Nonlinear Dynamics in the Life and Social Sciences}, W. Sulis and I. Trofimova (eds.), IOS Press, Amsterdam (2001).
\bibitem{Sulis-ct} W. Sulis,  NDPLS, 16(2), 113-136 (2012).
\bibitem{Sulis-o} W. Sulis, arXiv: 1204.2145.
\end{thebibliography}

\end{document}